\documentclass[12pt]{article}
\pdfoutput=1
\usepackage{typearea}
\typearea{12}
\usepackage{amsfonts,amsmath,amssymb}
\usepackage[pdftex]{graphicx}
\usepackage{bbm}
\usepackage{dsfont}

\newcommand{\pa}{\partial}
\newcommand{\nn}{\nonumber}
\newcommand{\Ord}{{\cal O}}

\makeatletter

\@addtoreset{equation}{section}
\makeatother

\def\href#1#2{#2}

\begin{document}

\begin{titlepage}

\begin{center}

\hfill 
\vskip 1.4in

\textbf{\Large Large Field Inflation Models From}\\[3mm]
\textbf{\Large Higher-Dimensional Gauge Theories}\\[14mm]

{Kazuyuki Furuuchi${\,}^a$ and Yoji Koyama${\,}^b$}
\vskip6mm
${}^a\,${\sl Manipal Centre for Natural Sciences, Manipal University}\\
{\sl Manipal, Karnataka 576104, India}\\[1mm]
${}^b\,${\sl Department of Physics, National Tsing-Hua University}\\
{\sl Hsinchu 30013, Taiwan R.O.C.}

\vskip8mm 
\end{center}
\begin{abstract}
Motivated by the recent detection of 
B-mode polarization of CMB by BICEP2
which is possibly of primordial origin,
we study large field inflation models
which can be obtained from higher-dimensional gauge theories.
The constraints from CMB observations 
on the gauge theory parameters are given, 
and their naturalness are discussed.
Among the models analyzed,
Dante's Inferno model turns out to be the
most preferred model in this framework.
\end{abstract}

\end{titlepage}

\section{Introduction}
Cosmic inflation 
\cite{Guth:1980zm,Linde:1981mu,%
Starobinsky:1980te,Sato:1980yn,Kazanas:1980tx,Albrecht:1982wi}
is a leading paradigm in the study of very early universe.
Inflation can explain not only the observed homogeneity and isotropy of the universe
over the super-horizon scale but also the tiny deviations from them 
\cite{Mukhanov:1981xt,Chibisov:1982nx,Guth:1982ec,Hawking:1982cz,Starobinsky:1982ee}.
The agreement between the general theoretical predictions 
of the standard slow-roll inflation and the recent
precise CMB measurements \cite{Ade:2013uln} is rather impressive.

Recently, another important clue from CMB observations came in.
BICEP2 team reported detection of B-mode polarization
at degree angular scales \cite{Ade:2014xna}.
While the important foreground analysis remains to be worked out in the future,
if the detected B-mode polarization turns out to be
of primordial origin,
it will have tremendous impacts on inflationary cosmology and
the understanding of our universe at its very beginning:
The tensor-to-scalar ratio 
fixes the energy scale at the time of inflation;
another important consequence of 
the large tensor-to-scalar ratio is that it
requires trans-Planckian inflaton field excursion 
via the 
Lyth bound \cite{Lyth:1996im}.
This poses a challenge for constructing viable inflation models,
since it is difficult to
protect the flatness of the potential from quantum corrections
over trans-Planckian field range in effective field theory framework. 
Thus the large tensor-to-scalar ratio
might require the knowledge of physics near the Planck scale.
However, this is not the only theoretical possibility:
Even if the effective field range of the inflaton is trans-Planckian,
field ranges in the defining theory can be sub-Planckian
\cite{Kim:2004rp,Silverstein:2008sg,McAllister:2008hb,%
Berg:2009tg,Tye:2014tja,Ben-Dayan:2014zsa,Kappl:2014lra,Long:2014dta,Bai:2014coa,%
Choi:2014rja,Kaloper:2008fb,Kaloper:2011jz,Marchesano:2014mla,%
Ashoorioon:2009wa,Harigaya:2014eta,McDonald:2014oza}.\footnote{%
This is a partial list of references on such models, 
we picked up papers 
whose interests are relatively close to that of the current paper.}
A subclass of this type of models 
which is specified below will be of our interest.

It has been known that
a gauge symmetry in
higher dimensions gives rise to an approximate 
shift symmetry in a four-dimensional scalar potential
\cite{Hosotani:1983xw,Hatanaka:1998yp},
and this mechanism was employed in 
\cite{ArkaniHamed:2003wu}
(see also \cite{Kaplan:2003aj})
to construct a version of
natural inflation \cite{Freese:1990rb}
(extra-natural inflation).
The original aim of \cite{ArkaniHamed:2003wu} 
was to construct a large field inflation model
(inflation model in which inflaton makes trans-Planckian field excursion)
within the framework of effective field theory.
But it was already noticed by the authors of \cite{ArkaniHamed:2003wu} 
that the embedding 
of extra-natural inflation to string theory was difficult,
and this point was further examined in \cite{Banks:2003sx}.
Then, it was suggested that the underlying reason for the difficulty was 
the extremely small gauge coupling which was required to explain the CMB data
in extra-natural inflation
\cite{ArkaniHamed:2006dz}.
The authors of \cite{ArkaniHamed:2006dz}
proposed that the tiny gauge coupling 
causes an obstacle for coupling the effective field theory to gravity.
It was motivated by the well-known argument against the existence of 
global symmetry in quantum gravity
based on processes involving black holes
(see \cite{Banks:2010zn} for recent discussions 
and references for earlier works).
When the gauge coupling is turned to zero,
the gauge symmetry is physically indistinguishable from a global symmetry.
If the limit to the zero gauge coupling is smooth, something 
must prevent the occurrence of the global symmetry.
The answer suggested in \cite{ArkaniHamed:2006dz} was that
when the gauge coupling becomes small, the
UV cut-off scale where the effective field theory breaks down
must be lowered.
More precisely, 
they proposed that there is an upper bound on the
UV cut-off scale $\Lambda$:
\begin{equation}
\Lambda \lesssim g M_P ,
\label{WGCintro}
\end{equation}
where $g$ is the gauge coupling
and $M_P$ is the four-dimensional Planck mass.
The authors of \cite{ArkaniHamed:2006dz} showed that the
bound (\ref{WGCintro})
follows from a conjecture that
there must be a particle whose mass is smaller
than its charge in certain unit
(Weak Gravity Conjecture, abbreviated as WGC below).
The basis of their arguments which lead to WGC are quite robust,
and in this paper we will take WGC seriously.
A brief review on WGC is given in appendix \ref{AppWGC}.

In this paper,
we examine large field inflation models
which can be obtained from
higher-dimensional gauge theories.
We restrict ourselves 
to one-form gauge fields in higher dimensions,
though these can appear from higher-form gauge fields
in even higher dimensions with smaller compactification size.
While in this paper we restrict ourselves to the simplest Abelian gauge groups,
it is straightforward to extend or embed our models 
to those with non-Abelian gauge groups.
Non-Abelian higher-form fields are known to be
theoretically quite involved (see e.g. \cite{Henneaux:1997ha}),
and our strategy of
first concentrating on one-form gauge fields 
may have an advantage in bypassing these theoretical complications
while still covering large portion of theory space.
Such one-form gauge fields are also essential ingredients
in the Standard Model of particle physics,
and it is natural to expect that one-form gauge fields 
will continue to be an essential part of
the new physics beyond the Standard Model.
These constitute our basic motivations to consider 
one-form gauge theories in higher dimensions.

We are particularly interested in the consequences of WGC,
and will assume that it is correct.\footnote{%
Another possibility would be that WGC does not always hold, 
but holds in the dominant majority of string vacua.
While this is an interesting theoretical possibility,
it is not relevant for the discussion of naturalness below
as long as it is extremely likely to be in a vacuum in which WGC holds.}
Thus the original extra-natural inflation will be excluded
from our study.\footnote{%
There is a possibility that WGC
completely excludes natural parameter space for effective field theory.
In this case, one may respect the constraints 
from WGC
and accept the unnatural values of the parameters.
See \cite{Cheung:2014vva} for an argument 
on an example in particle physics model.
In this paper we will be interested in
natural parameter space allowed by WGC.}
This naturally lead us to consider models of the type mentioned above:
Those in which the field ranges in the defining theory are sub-Planckian
but the inflaton effectively travels trans-Planckian field range.
As higher-dimensional gauge theories reduce to
so-called axion models,
we examined all the major axionic inflation models of the above mentioned type
so far known to us,
at least in their simplest form.
These include:
Single-field Axion Monodromy model (AM) \cite{Silverstein:2008sg,McAllister:2008hb},
Dante's Inferno model (DI) \cite{Berg:2009tg},
Axion Alignment model (AA) \cite{Kim:2004rp,Kappl:2014lra,Long:2014dta,Bai:2014coa}
and Axion Hierarchy model (AH) \cite{Tye:2014tja,Ben-Dayan:2014zsa}.
We will examine the constraints from CMB data on gauge theory parameters
and discuss their naturalness in the effective field theory framework.
However, for the tensor-to-scalar ratio,
the above mentioned BICEP2 result does not give conclusive value
due to the uncertainty in the foreground \cite{Flauger:2014qra,Adam:2014bub}.
In this paper, we would like to explore the possibility
that the large tensor-to-scalar ratio is real
considering its impact if it turns out to be the case.
We choose $r=0.16$ at the pivot scale as a reference value \cite{Audren:2014cea},
but this should be taken as an assumption at this moment.

Table \ref{prior}
summarizes the expected parameter ranges in our models.
While we will not go into full Bayesian model comparison 
(see e.g. \cite{Trotta:2008qt,Bayesian}),
in principle we can go through it, 
and in that case
our prior can be built based on Table \ref{prior}.
In Table \ref{prior}, $g$ stands for four-dimensional gauge coupling
which is obtained from higher-dimensional gauge theory as
\begin{equation}
\frac{1}{g^2} = \frac{2\pi L}{g_5^2} ,
\label{ig4g5}
\end{equation}
where $g_5$ is the five-dimensional gauge coupling and
$L$ is the compactification radius of the fifth dimension.
$g_5^2$ has dimension of length
which can be independent from the compactification radius.
A priori, we do not have knowledge of their corresponding energy scales
besides the upper bound by the Planck scale
and lower bound from high energy experiments like LHC.
Therefore, the log-flat prior would be appropriate for $g$ and $L$,
if we were to proceed to Bayesian model comparison.
The lower bound in $g$ in Table \ref{prior} is imposed by WGC,
while the upper bound comes from applicability of perturbation theory.
The expected value of charges is shown in Table \ref{prior} 
in unit of the minimal charge in the model.
It reflects the theoretical belief 
of the current authors 
that extraordinary large charge is 
unlikely or rare in nature.

Table \ref{gposterior}-\ref{mnposterior}
show the allowed parameter ranges
after taking into account CMB data and assuming $r=0.16$.
Strictly speaking,
it is more appropriate to show the allowed parameter range in
multi-dimensional parameter space, as the 
allowed range for one parameter depends on other parameters in general.
However, even in the current simplified analysis,
one immediately notices that 
somewhat unusual parameter ranges appear
in Table \ref{mnposterior}:\footnote{As can be seen 
from the derivation of Table \ref{mnposterior}
in the main body, this conclusion does not depend on other parameters.}
AA
and 
AH
have at least one charge which is more than $\Ord(100)$
in unit of minimal charge in the model.
Although theories with such a large charge number have been considered,
(e.g. see \cite{Shiu:2013wxa} for the so-called milli-charged dark matter,
where an issue related to WGC is discussed),
such theories look somewhat artificial.
This view of the current authors had been reflected
in the expected charge number in Table \ref{prior}.
On the other hand, the charge of 
AM
is in a natural range, but this model has
its own naturalness issue which will be explained in section \ref{secSA}.
Charges in DI are in the expected range 
given in Table \ref{prior}.
From these analysis, one immediately sees
that DI is preferred among the models considered.

\begin{table}
\begin{center}
  \begin{tabular}{  | c | c | c | }
    \hline
	Gauge couplings & Compactification radius & Charges  \\ 
		\hline
  $ -\log_{10} [(LM_P)^2] \lesssim \log_{10} [g^2] \lesssim 0$ & $\log_{10} [1/(L\,{\rm GeV})] \sim 3-17$ & $n \sim \Ord(1)$\\ 
    \hline
  \end{tabular}
\caption{Expected parameter ranges from higher-dimensional gauge theory.
$g$ is the gauge coupling in four-dimension.
$L$ is the compactification radius of the fifth dimension.
$n$ represents charge of a matter
measured in unit of the minimal charge in the model.}
\end{center}
\label{prior}
\end{table}

\begin{table}
\begin{center}
\begin{tabular}{|c|c|}
\hline
Model & Gauge coupling(s) \\
\hline
AM & $-8 \lesssim \log_{10} [g^2] \lesssim 0$\\
\hline
DI & $ -1 \lesssim \log_{10} [g_A^2] \lesssim 0$, $-3 \lesssim \log_{10} [g_B^2] \lesssim -2$ \\
\hline
AA & $-10 \lesssim \log_{10} [g_A^2], \log_{10} [g_B^2] \lesssim -4$ \\
\hline
AH & $-10 \lesssim \log_{10} [g_A^2] \lesssim -4$,
$-10 \lesssim \log_{10} [g_B^2] \lesssim 0$  \\
\hline
\end{tabular}
\caption{Constraints on gauge couplings 
after taking into account CMB data with the assumption $r=0.16$.}
\label{gposterior}
\end{center}
\end{table}
\begin{table}
\begin{center}
\begin{tabular}{|c|c|}
\hline
Model & Compactification radius \\
\hline
AM & $\log_{10} [1/(L\, {\rm GeV})] \sim 14-16$ \\
\hline
DI & $\log_{10} [1/(L\, {\rm GeV})] \sim 17$ \\
\hline
AA & $\log_{10} [1/(L\, {\rm GeV})] \sim 14-17$ \\
\hline
AH & $\log_{10} [1/(L\, {\rm GeV})] \sim 14-17$ \\
\hline
\end{tabular}
\caption{Constraints on compactification radius 
after taking into account CMB data with the assumption $r=0.16$.}
\label{Lposterior}
\end{center}
\end{table}
\begin{table}
\begin{center}
\begin{tabular}{|c|c|}
\hline
Model & Charge(s) \\
\hline
AM & $\Ord(1)$\\
\hline
DI & $\Ord(1)$ \\
\hline
AA & max$(|m_1,m_2|) \gtrsim \Ord(100)$ \\
\hline
AH & $m_1 \gtrsim \Ord(100)$\\
\hline
\end{tabular}
\caption{Constraints on charges
after taking into account CMB data with the assumption $r=0.16$.}
\label{mnposterior}
\end{center}
\end{table}

The organization of the rest of the paper is as follows.
We start with 
single-field Axion Monodromy model 
in section \ref{secSA}.
In section \ref{secDI} we study Dante's Inferno model. 
In section \ref{secAlign} 
Axion Alignment model 
and 
Axion Hierarchy model 
are studied.\footnote{In \cite{Bai:2014coa} aligned natural inflation
from higher-dimensional gauge theory similar to ours was studied,
but the four-dimensional WGC was not imposed.}
For each model we obtain it from higher-dimensional gauge theory,
study the constraints from the CMB observations to
the parameters of the gauge theory and
discuss naturalness of the parameters.
We summarize with discussions on future directions in section \ref{secSummary}.

\section{Single-Field Axion Monodromy}\label{secSA}
We begin with single-field axion monodromy inflation 
\cite{Silverstein:2008sg,McAllister:2008hb}.
The relevant inflaton potential is of the form
\begin{equation}
V(A) 
= \frac{1}{2} m^2 A^2 
 + \Lambda^4 \left(1 - \cos \left(\frac{A}{f} \right) \right).
\label{pot1}
\end{equation}
The potential (\ref{pot1}) can be obtained from a 
five-dimensional gauge theory with an action\footnote{%
We chose the massless charged fermion for an illustrative purpose.
We can introduce mass term for the fermion
or include charged massive scalars in a similar way.}
\begin{equation}
S =
\int d^5 x
\Bigl[
-\frac{1}{4} F_{MN} F^{MN}
-\frac{1}{2} m^2 
\left(A_M - g_5\pa_M \theta \right)^2
+ (\mbox{matters})
\Bigr].
\end{equation}
We introduced the Stueckelberg mass term
which gives rise to the quadratic potential in (\ref{pot1}).%
\footnote{Massive gauge fields 
can arise via the Higgs mechanism.
However,
the expectation value of the 
radial component of the Higgs field,
which determines the mass of the gauge field, 
is affected by the large 
inflaton expectation value,
as the inflaton originates from gauge field in the current model
and couples to the Higgs field as such.
Then the current analysis does not apply.
For a recent review
on the use of Stueckelberg fields in axion monodromy inflations
in string theory, see \cite{McAllister:2014mpa}.}
We take the gauge group to be compact $U(1)$.\footnote{%
It has been argued that in models which can be 
consistently coupled to quantum gravity,
all the continuous gauge symmetries are compact
\cite{Banks:2010zn}.}
Then, the Stueckelberg field
$\theta$ is an angular variable with the identification
\begin{equation}
\theta \sim \theta + \frac{2 \pi}{g_5}.
\label{thang}
\end{equation}
This allows $\theta$ to have 
a winding mode:
\begin{equation}
\theta(x,x^5) =  \frac{x^5}{g_5 L} w + \sum_n \theta_n(x) e^{i\frac{n}{L}x^5}
\label{wind}
\end{equation}
Here, $x$ are coordinates in visible large space-time dimensions,
and $x^5$ is the coordinate of the fifth direction compactified on a circle
with radius $L$.
The winding number $w$ is an integer.
If one takes into account all the winding sectors,
the spectrum of the model is invariant under the 
shift of $A$ by $2\pi f$,
while starting from a sector with 
given winding number the shift leads to the monodromy property
\cite{Silverstein:2008sg,McAllister:2008hb}.
At one-loop, the following potential is generated:
\begin{equation}
V(A) 
= \frac{1}{2} m^2 \left(A - {2\pi f w} \right)^2 
 + \Lambda^4 \left(1 - \cos \left(\frac{A}{f} \right) \right).
\label{pot2}
\end{equation}
See appendix \ref{AppOne} for the outline of the 
calculation of the one-loop effective potential.
For a sector with a given winding number,
by redefining $A$ by a constant shift 
one obtains (\ref{pot1}).
The inflaton field $A$ in the potential (\ref{pot1})
is the zero-mode of the gauge field:
\begin{equation}
A \equiv A_{5(0)}.
\end{equation}
The parameters of the axion monodromy model
(\ref{pot1}) are related to the parameters of
the higher-dimensional gauge theory as follows:
\begin{equation}
f = \frac{1}{g (2\pi L)}, \quad
\Lambda^4 = \frac{c}{\pi^2(2\pi L)^4}, \quad c \sim \Ord(1),
\label{gparas}
\end{equation}
where $g$ is the four-dimensional gauge coupling
which is related to the five-dimensional gauge coupling $g_5$ as
\begin{equation}
g = \frac{g_5}{\sqrt{2 \pi L}}.
\label{g4g5}
\end{equation}
The constant $c$ in (\ref{gparas})
depends on the matter contents charged under the gauge group.
In (\ref{gparas}) we have assumed that
both the number of the matter fields and their charges are of order one,
which we think natural.

If one considers all possible winding numbers of $\theta$,
the whole theory is invariant under the shift 
$A \rightarrow A + 2\pi f$.
Thus the field $A$ takes values on a circle with radius $f$.
Starting from a given winding number sector, 
the quadratic potential reveals the phenomenon
of monodromy:
The potential energy does not return the same under
the shift of $A$ by $2\pi f$.
Thus one can effectively achieve trans-Planckian field excursion of $A$
even if the original period of $A$ was below the reduced Planck scale
$M_P = 2.4 \times 10^{18}$ GeV,
by going round the circle several times.
This is an important feature of the model, 
because
examples in string theory so far constructed
and WGC
suggest $2\pi f \lesssim M_P$
for an axion decay constant $f$,
which 
forbids trans-Planckian field excursion of
the axion if there were no monodromy
(see appendix \ref{AppWGC} for
the assertions of WGC 
we adopt in this paper).

When 
the slope of the sinusoidal potential
is much smaller than that of the mass term
in (\ref{pot1}),
the model effectively reduces to 
chaotic inflation.\footnote{%
See \cite{Flauger:2009ab,Meerburg:2014bpa} for the case 
in which the sinusoidal potential is not totally negligible.
From appendix A of \cite{Meerburg:2014bpa}
one can show that the effect of the sinusoidal potential
is proportional to $L^{-3}$ and thus quickly suppressed
as one moves away from the bound in (\ref{invL}).}
This condition is written as
\begin{equation}
\frac{\Lambda^4}{f} \ll m^2 A_\ast,
\label{mtdom}
\end{equation}
where $A_\ast$ is the value of $A$ when
the pivot scale exited the horizon.
Using (\ref{gparas}), this condition becomes
\begin{equation}
\frac{3g}{\pi^2}\frac{1}{(2\pi L)^3}
\ll m^2 A_\ast,
\label{mtdom2}
\end{equation}
or
\begin{equation}
\frac{1}{L} < 2\pi \left( \frac{\pi^2}{3g} m^2 A_\ast \right)^{1/3}.
\label{invL}
\end{equation}
We review the constraints from CMB observations on chaotic inflation
in appendix \ref{appchao}.
Putting the values of
$m^2$ and $A_\ast$
given in (\ref{m}) and (\ref{phipiv})
for $r=0.16$ and $N_\ast\simeq 50$,
we obtain
\begin{equation}
\frac{1}{L} < g^{-1/3} \times 3.2 \times 10^{16} \,\rm{GeV}.
\label{invLval1}
\end{equation}
Note that
the energy scale of the compactification 
should not be smaller than the Hubble scale during inflation,
otherwise the use of the four-dimensional Einstein equation
is not justified.
From (\ref{H}), this gives
\begin{equation}
\frac{1}{L} > 1.0 \times 10^{14} \,\rm{GeV}.
\label{invLval2}
\end{equation}

If there were no sinusoidal potential,
when one takes $m^2$ to zero the shift symmetry
$A \rightarrow A + c$ ($c$: constant) recovers.
Thus small $m^2$ is natural 
in the sense of 't Hooft \cite{'tHooft:1979bh}.
In order for the inflaton to achieve
trans-Planckian field excursion,
this shift symmetry must be a good symmetry at the Planck scale.
Whether this is the case or not 
is a problem beyond the scope of the higher-dimensional gauge theory,
which is an effective field theory.
One needs to work in a theory of quantum gravity to study this issue.
In other words, while the whole theory is invariant under the shift of 
the field $A$ by $2\pi f$,
starting from a given winding number
the potential of $A$ is not periodic.
And the large $A$ behavior of the non-periodic part of the potential
has the usual UV issue of effective field theory.

\section{Dante's Inferno}\label{secDI}

Next we study
Dante's Inferno model \cite{Berg:2009tg},
which is a two-axion model with 
the following potential:
\begin{equation}
V(A,B) 
= \frac{1}{2} m_A^2 A^2 
 + \Lambda^4 \left(1 - \cos \left(\frac{A}{f_A} - \frac{B}{f_B} \right) \right).
\label{potDI}
\end{equation}
The potential (\ref{potDI})
can be obtained 
from a gauge theory in higher dimensions
with the action
\begin{align}
S =
\int d^5 x&
\Bigl[
-\frac{1}{4} F_{MN}^{A} F^{AMN}
-\frac{1}{2} m_A^2 
\left(A_M - g_{A5} \pa_M \theta \right)^2
-\frac{1}{4} F_{MN}^{B} F^{BMN} \nn \\
&- i \bar{\psi} \gamma^M \left( \pa_M + i g_{A5} A_M - i g_{B5} B_M \right) \psi 
\Bigr].
\end{align}
We consider the case where both of the gauge groups are compact $U(1)$,
which we refer to as $U_{A}(1)$ and $U_{B}(1)$.
Here, as an illustration, we consider fermionic matter,
but the case with bosonic matters can be studied in essentially the same way.
The one-loop effective potential of this model produces
the second term in (\ref{potDI}) with
\begin{equation}
{f_{A}} = \frac{1}{g_{A} (2\pi L)}, \quad 
{f_{B}} = \frac{1}{g_{B} (2\pi L)} ,
\label{fs}
\end{equation}
and
\begin{equation}
\Lambda^4 \simeq \frac{3}{\pi^2}\frac{1}{(2\pi L)^4}.
\label{Lambda}
\end{equation}
Here, $g_A$ and $g_B$ are four-dimensional gauge couplings 
which are related to the five-dimensional gauge couplings 
$g_{A5}$ and $g_{B5}$ as
\begin{equation}
g_A = \frac{g_{A5}}{\sqrt{2\pi L}}, \quad
g_B = \frac{g_{B5}}{\sqrt{2\pi L}} .
\label{gAB}
\end{equation}
It is convenient to rotate the fields as
\begin{equation}
\left(
\begin{array}{c}
	\tilde{B}\\
	\tilde{A}
\end{array}
\right)
=
\left(
\begin{array}{cc}
	\cos \gamma & \sin \gamma \\
	-\sin \gamma & \cos \gamma
\end{array}
\right)
\left(
\begin{array}{c}
	B\\
	A
\end{array}
\right),
\label{rot}
\end{equation}
where
\begin{equation}
\sin \gamma = \frac{f_A}{\sqrt{f_A^2 + f_B^2}},
\quad
\cos \gamma = \frac{f_B}{\sqrt{f_A^2 + f_B^2}}.
\label{gamma}
\end{equation}
Then the potential (\ref{potDI}) takes the form
\begin{equation}
V(\tilde{A},\tilde{B})
=
\frac{m_A^2}{2}
\left(
\tilde{A} \cos \gamma + \tilde{B} \sin \gamma
\right)^2
+
\Lambda^4
\left(
1 - \cos \frac{\tilde{A}}{f}
\right),
\label{pott}
\end{equation}
where
\begin{equation}
f \equiv \frac{f_A f_B}{\sqrt{f_A^2 + f_B^2}}.
\label{f}
\end{equation}

In this model, the regime of interest is\footnote{%
Be aware of the difference between 
(\ref{mtdom}) and (\ref{slope}).}
\begin{align}
2\pi f_A &\ll 2\pi f_B \lesssim M_P ,\label{fAfB}\\
\frac{\Lambda^4}{f} &\gg m_A^2 A_{in}, \label{slope}
\end{align}
where $A_{in}$ is the initial condition
set at the beginning of the observable inflation
and we require it to be in the range
$f \ll A_{in} < M_P$.
Notice that the condition (\ref{fAfB})
implies in the leading order in $f_A/f_B$
\begin{equation}
\cos \gamma \simeq 1, \quad \sin \gamma \simeq \frac{f_A}{f_B},
\quad f \simeq f_A.
\label{implied}
\end{equation}
We require that the
excitation in $\tilde{A}$ direction 
is much heavier than the Hubble scale during inflation
so that they can be safely integrated out:
\begin{equation}
\frac{\partial^2}{\partial \tilde{A}^2} V (\tilde{A},\tilde{B})
> H^2.
\label{ddVH}
\end{equation}
From (\ref{slope}) and $f\ll A_{in}$
this reads
\begin{equation}
\frac{3g_A^2}{\pi^2 L^2} > H^2.
\label{gALH}
\end{equation}
After integrating out $\tilde{A}$,
we obtain the following effective potential for $\tilde{B}$
which we rewrite as $\phi \equiv \tilde{B}$
\cite{Berg:2009tg}:
\begin{equation}
V_{{\rm eff}}(\phi)
=
\frac{m^2}{2} \phi^2,
\quad
m\equiv \frac{f_A}{f_B} m_A,
\label{poteff}
\end{equation}
to leading order in $f_A/f_B$.
Thus Dante's Inferno model
effectively reduces to
chaotic inflation, with $\phi$ being the inflaton.
The constraints 
from CMB observations on chaotic inflation
are summarized in appendix \ref{appchao}.
Using these inputs, 
now we examine the CMB constraints on the
parameters of the higher-dimensional gauge theory.
We will take the number of e-fold $N_\ast \simeq 50$ and 
the tensor-to-scalar ratio $r=0.16$
(see appendix \ref{AppCMB} for the detail and the notations
used below).
From (\ref{fs}),
the condition (\ref{fAfB}) reads in terms of gauge theory parameters as
\begin{equation}
g_A \gg g_B,
\label{A}
\end{equation}
and
\begin{equation}
\frac{1}{g_B (2 \pi L)} \lesssim {M_P}.
\label{B}
\end{equation}

Chaotic inflation is a large field inflation model
in which the inflaton travels trans-Planckian field distance
$\Delta \phi \equiv \phi_\ast - \phi_e \simeq 14 M_P$,
see (\ref{phipiv}).
However, the original fields in the current model, $A$ and $B$
(which were the zero-modes of the higher-dimensional gauge theory),
do not need to make trans-Planckian field excursion.

Regarding the field $A$, its initial value $A_{in}$ is restricted as
\begin{equation}
A_{in} \simeq \frac{f_A}{\sqrt{f_A^2 + f_B^2}} \phi_\ast
\simeq \frac{f_A}{f_B} \times 14 M_P.
\label{Ain}
\end{equation}
Thus $A_{in}$ is sub-Planckian if
\begin{equation}
f_A < \frac{1}{14} f_B .
\label{fAfBb}
\end{equation}
From (\ref{fs}), in terms of gauge couplings (\ref{fAfBb}) amounts to
\begin{equation}
g_A > 14 g_B. 
\label{gAgB}
\end{equation}
This condition should be compared with (\ref{A}).
On the other hand, field $B$ is periodic and its field range $2\pi f_B$ is 
bounded from above by $M_P$, as noted in (\ref{fAfB}).

\begin{figure}[htbp]
\centering
\includegraphics[width=5in]{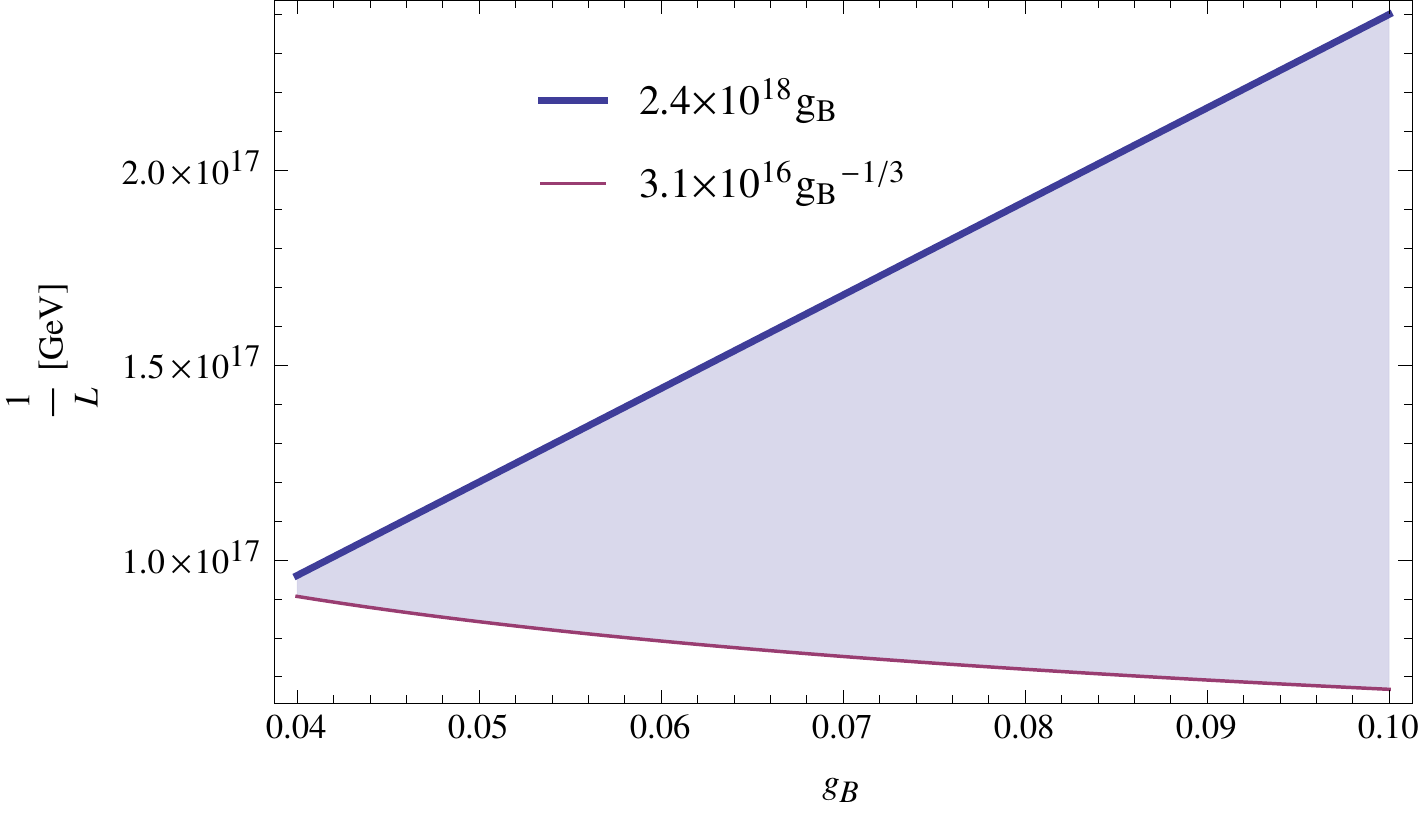} 
\caption{\label{fig:gL} Allowed range of $L$ as a function of $g_B$.}
\end{figure}

There is also a lower bound on the inverse compactification radius.
Using (\ref{poteff}) and (\ref{Ain}), 
the condition (\ref{slope}) can be rewritten as
\begin{equation}
\frac{3}{\pi^2 (2\pi L)^4} \gg 
f_A \left(m \frac{f_B}{f_A} \right)^2
\frac{f_A}{f_B} \times 14 M_P .
\label{BD}
\end{equation}
Using (\ref{fs}) and
putting the value of $m$ in (\ref{m}),
we obtain
\begin{equation}
\frac{g_B^{1/3}}{L} >
3.2 \times 10^{16} \,{\rm GeV}.
\label{gBL3val}
\end{equation}
Together with (\ref{B})
we have
\begin{equation}
g_B^{-1/3} \times 3.2 \times 10^{16} \,{\rm GeV}
<
\frac{1}{L} 
\lesssim
g_B 
\times 2.4 \times 10^{18} \,{\rm GeV}.
\label{Lrange}
\end{equation}
(\ref{Lrange}) immediately implies $g_B \gtrsim 0.04$.
On the other hand, in order for our one-loop effective potential to be valid,
the gauge coupling should not be large, $g_A \lesssim \Ord(1)$. 
Together with (\ref{gAgB}), we have 
\begin{equation}
0.04 \lesssim g_B \lesssim \Ord(0.1).
\label{gBrange}
\end{equation}
For $g_B =0.04$ we have
\begin{equation}
9.2 \times 10^{16} \,{\rm GeV}
<
\frac{1}{L} 
\lesssim
9.6 \times 10^{16} \,{\rm GeV} ,
\label{Lrangeval}
\end{equation}
while for $g_B=0.1$ we have
\begin{equation}
6.8 \times 10^{16} \,{\rm GeV}
<
\frac{1}{L} 
\lesssim
2.4 \times 10^{17}\, {\rm GeV}.
\label{Lrange2}
\end{equation}
See Fig.~\ref{fig:gL} for the values of $g_B$ in between.
We observe that the allowed values of the gauge couplings
and the compactification radius of the gauge theory
are rather restricted,
which will be advantageous for the model to be predictive.
Note that the above compactification scales are high enough 
so that the use of the four-dimensional Einstein gravity
is justified, $1/L \gg H \sim 10^{14}$ GeV (see (\ref{H})).

For completeness, we check that 
(\ref{ddVH}) is satisfied.
It gives
\begin{equation}
\frac{g_A}{2\pi L} \gtrsim \frac{\pi}{\sqrt{3}} H .
\label{gAL}
\end{equation}
Putting the value from 
appendix \ref{AppCMB} (\ref{H})
we obtain 
\begin{equation}
\frac{g_A}{2\pi L} \gtrsim 2 \times 10^{14}\, {\rm GeV}.
\label{gALval}
\end{equation}
This is readily satisfied for the above values of $g_A$ and $L$.

Now we turn to another feature of the model
which could be potentially constrained by CMB data.
The shift symmetry allows
the following axionic coupling to gauge fields:
\begin{equation}
S_{AC} 
= 
\int d^4x \,
\frac{\alpha_i \sigma_i}{4f_i} F_{\mu\nu} \tilde{F}^{\mu\nu} ,
\label{AC}
\end{equation}
where $\sigma_i$ is an axion,
$f_i$ is its decay constant and $\alpha_i$ is a constant parameter.
$i$ labels axions when there are more than one,
in the current case $i$ labels the field $A$ and the field $B$
(we just label them as $i=A$ and $i=B$, respectively).
How the coupling (\ref{AC}) arises from higher-dimensional gauge theory
is explained in appendix \ref{appAC}.
Contributions to CMB power spectrum, 
non-Gaussianity and primordial gravitational waves 
through this coupling have been studied in
\cite{Anber:2006xt,Barnaby:2011qe,Meerburg:2012id,Linde:2012bt,Ferreira:2014zia}.
These effects 
are mainly controlled by the following parameter:
\begin{equation}
\xi_i \equiv \frac{\alpha_i \dot{\sigma}_i}{2f_i H} .
\label{xi}
\end{equation}
The current observational bound is given as
\cite{Meerburg:2012id,Ferreira:2014zia}
\begin{equation}
\xi_i \lesssim 3 .
\label{xibound}
\end{equation}

To obtain $\xi_i$ $(i=A,B)$ in (\ref{xi}),
we first need to know the time derivatives of fields $A$ and $B$.
In $\tilde{A}$ direction, we had
\begin{equation}
\dot{\tilde{A}} = 0.
\label{dtildA}
\end{equation}
On the other hand, $\tilde{B}$ is the inflaton
which slowly rolls down the potential.
From (\ref{epsr}) we estimate
\begin{equation}
\frac{\dot{\tilde{B}}^2}{H^2 M_P^2} 
\sim \frac{M_P^2}{2}\left(\frac{V'(\tilde{B})}{V(\tilde{B})}\right)^2
\sim 0.01 .
\label{eB}
\end{equation}
From (\ref{dtildA}) and (\ref{eB}) 
we can estimate $\dot{A}$ and $\dot{B}$ through
\begin{align}
\dot{A} 
&= \sin \gamma \dot{\tilde{B}} + \cos \gamma \dot{\tilde{A}} 
\sim \frac{f_A}{f_B} \dot{\tilde{B}},
\label{dotA}\\
\dot{B}
&=
\cos \gamma \dot{\tilde{B}} - \sin \gamma \dot{\tilde{A}}
\sim
\dot{\tilde{B}}.
\label{dotB}
\end{align}
On the other hand, $\alpha_i$ $(i=A,B)$
can be obtained as in (\ref{alpha}):
\begin{equation}
\alpha_A = \frac{g_A^2 k_A}{4\pi^2},\quad
\alpha_B = \frac{g_B^2 k_B}{4\pi^2}
\label{alphas}
\end{equation}
Putting (\ref{eB}), (\ref{dotA}), (\ref{dotB})
and (\ref{alphas}) into the definition (\ref{xi}),
we arrive at
\begin{align}
\xi_A 
&\lesssim
\frac{g_A^3 k_A}{4\pi} LM_P \times \frac{0.1}{14},
\label{xiA}\\
\xi_B 
&\sim
\frac{g_B^3 k_B}{4\pi} LM_P \times 0.1,
\label{xiB}
\end{align}
In deriving (\ref{xiA}) we have used (\ref{fAfBb}).
As we have assumed $g_A \lesssim \Ord(1)$,
by putting
(\ref{fAfBb}) and
$L \sim \Ord (10^{17})$ GeV,
we obtain
\begin{equation}
\xi_A \lesssim k_A \times \Ord(10^{-2}).
\label{xiAb}
\end{equation}
On the other hand, 
from
$g_B \lesssim \Ord(0.1)$ in (\ref{gBrange}), 
we obtain
\begin{equation}
\xi_B \lesssim k_B \times \Ord (10^{-3}).
\label{xiBb}
\end{equation}
As argued in appendix \ref{appAC},
we expect $k_A, k_B \sim \Ord(1-10)$.
In this case the observational bound
$\xi_i \lesssim 3$ is satisfied for $i=A,B$.

\section{Axion Alignment and Axion Hierarchy}\label{secAlign}

In this section we study
aligned axion inflation
\cite{Kim:2004rp,Kappl:2014lra,Long:2014dta} 
and 
hierarchical axion inflation 
\cite{Tye:2014tja,Ben-Dayan:2014zsa}
from higher-dimensional gauge theory perspective.
Both models can be described by the potential
of the form
\begin{equation}
V(A,B) 
=  
\Lambda_1^4  
\left( 1 - \cos \left(\frac{m_1}{f_A} A + \frac{n_1}{f_B} B \right) \right) 
 + 
\Lambda_2^4 \left( 1 - \cos \left( \frac{m_2}{f_A} A + \frac{n_2}{f_B}B \right) \right).
\label{poth}
\end{equation}
Upon field rotation
\begin{equation}
\left(
\begin{array}{c}
	\phi_s \\
	\phi_l
\end{array}
\right)
=
\left(
\begin{array}{cc}
	\cos \zeta & \sin \zeta\\
	-\sin \zeta & \cos \zeta
\end{array}
\right)
\left(
\begin{array}{c}
	A \\
	B
\end{array}
\right),
\label{roth}
\end{equation}
with
\begin{equation}
\cos \zeta = \frac{f_s}{f_A} m_1, \quad 
\sin \zeta = \frac{f_s}{f_B} n_1, 
\end{equation}
\begin{equation}
{f_s} 
=
\frac{1}{\sqrt{\frac{m_1^2}{f_A^2}+\frac{n_1^2}{f_B^2}}},
\label{zeta}
\end{equation}
the potential (\ref{poth}) takes the form
\begin{equation}
V(\phi_s,\phi_l) 
=  
\Lambda_1^4  
\left( 1 - \cos \left(\frac{\phi_s}{f_s}  \right) \right) 
 + 
\Lambda_2^4 \left( 1 - \cos \left( \frac{\phi_s}{f_s'}  + \frac{\phi_l}{f_l} \right) \right),
\label{poth2}
\end{equation}
where
\begin{equation}
f_l 
=
\frac{\sqrt{m_1^2 f_B^2 + n_1^2 f_A^2}}{m_1n_2 - m_2n_1},
\quad
f_s'=
\frac{1}{f_s\left(\frac{m_1m_2}{f_A^2}+\frac{n_1n_2}{f_B^2}\right)}.
\label{fl}
\end{equation}

The potential (\ref{poth})
can be obtained from a
higher-dimensional gauge theory
with following action:
\begin{align}
S =
\int d^5 x&
\Bigl[
-\frac{1}{4} F_{MN}^{A} F^{AMN}
-\frac{1}{4} F_{MN}^{B} F^{BMN} \\
&- i \bar{\psi} \gamma^M (\pa_M + i g_{A5} m_1 A_M + i g_{B5} n_1 B_M) \psi
- i \bar{\chi} \gamma^M (\pa_M + i g_{A5} m_2 A_M - i g_{B5} n_2 B_M) \chi
\Bigr]. \nn
\end{align}
The parameters in the potential (\ref{poth}) and 
the higher-dimensional gauge theory are related as
\begin{equation}
f_A = \frac{1}{g_A (2\pi L)}, \quad f_B = \frac{1}{g_B (2\pi L)},
\label{fAfBalign}
\end{equation}
where $g_A$ and $g_B$ are four-dimensional gauge couplings
\begin{equation}
g_A \equiv \frac{g_{A5}}{\sqrt{2\pi L}}, \quad
g_B \equiv \frac{g_{B5}}{\sqrt{2\pi L}}.
\label{g4s}
\end{equation}
Anticipating 
UV completions such as string theory,
it is natural that charges are quantized with respect to the unit charge.
Thus we assume $m_1$, $m_2$, $n_1$, $n_2$ are all integers.\footnote{%
As we have assumed that the gauge groups are compact $U(1)$,
charges are quantized.
Here we made a stronger assumption that
charges are all integer multiples of the minimal charge in the theory.
This can be regarded as for simplicity, the result does not change
qualitatively unless one assumes highly exotic charge spectrum.}

Aligned
axion inflation 
is obtained in the regime
\begin{equation}
|m_1n_2 - m_2n_1| \ll |m_1|, |n_1|.
\label{aligned}
\end{equation}
In this regime 
one obtains $|f_l| \gg f_A, f_B$ from (\ref{fl}).
Notice that $|f_l|$ is at largest the order of $\max (|m_1| f_B, |n_1| f_A)$.
On the other hand,
as explained in appendix \ref{appnatural},
$r\simeq 0.16$ requires
$|f_l| \gtrsim 20 M_P$.
Since from WGC we have
$2\pi f_A, 2\pi f_B \lesssim M_P$,
this requires $\max(|m_1|,|n_1|) \gtrsim 20\times 2\pi$.
A matter with such a large charge seems to us quite unnatural,
considering that 
the energy scale under consideration
is rather high ($H \sim 10^{14}$ GeV).

Next we turn to the hierarchical axion inflation
in higher-dimensional gauge theory.
This model corresponds to taking
$n_2 =0$ in (\ref{poth}).
Then (\ref{fl}) reduces to 
\begin{equation}
|f_l| = \frac{\sqrt{m_1^2 f_B^2 + n_1^2 f_A^2}}{|n_1m_2|}.
\label{fl2}
\end{equation}
One further requires a hierarchy
\begin{equation}
 \left|\frac{f_A}{m_1}\right| \ll \frac{f_A}{|m_2|} , \frac{f_B}{|n_1|} .
\label{fhier}
\end{equation}
Then (\ref{fl2}) can be approximated as
\begin{equation}
|f_l| \simeq \left|\frac{m_1}{n_1 m_2}\right| f_B .
\label{fl3}
\end{equation}
From WGC we have
$2\pi f_B \lesssim M_P$, thus
$|f_l| \gtrsim 20 M_P$
requires
$|m_1| \gtrsim 20 |n_1 m_2| \times 2\pi$.
Such a large hierarchy between the charges
in the same gauge group seems quite unnatural.\footnote{%
The upper bound of the gauge coupling in Table \ref{gposterior}
for AA and AH were obtained by requiring applicability of
perturbation theory with these large charge number:
In order for the perturbation theory to be appropriate,
we need $g n \sim \Ord(1)$,
where $g$ is the gauge coupling and $n$ 
is the maximal charge in the model.}

\section{Summary and Discussions}\label{secSummary}

In this paper we studied large field inflation models
which can be obtained from higher-dimensional gauge theories.
We accept WGC as our working hypothesis, 
and studied the constraints from CMB data on the gauge theory parameters.
We consider the case with large tensor-to-scalar ratio,
and used $r=0.16$ as a reference value.
We found that the allowed range of gauge theory parameters are quite constrained.
Among the models 
studied in this paper,
Dante's Inferno model
appears as the most preferred model. 
The allowed values of the gauge couplings and the compactification radius
turned out to be quite restricted but fell within a natural range,
making the model attractive for being predictive.
Single-field axion monodromy model
leaves the problem that whether the 
shift symmetry is a good symmetry or not 
to its UV completion.
Axion alignment model and axion hierarchy model
require large hierarchy among
charges in the same gauge group,
which makes the models rather unnatural.

The allowed values of gauge couplings in Dante's Inferno model are 
in the range $0.04 - \Ord(1)$.
This is in contrast to the extremely small gauge coupling
$\lesssim \Ord(10^{-3})$ required for extra-natural inflation
\cite{ArkaniHamed:2003wu,Furuuchi:2013ila}.
The above values of gauge couplings 
for Dante's Inferno model
would be large enough to have interesting
consequences in cosmological history or particle physics experiments
in model dependent ways, which will be interesting to investigate.
In particular, since gauge symmetry is a basic ingredient 
of the Standard Model of particle physics,
it is natural to expect that 
the higher-dimensional gauge theories responsible for inflation
are also relevant for the new physics beyond the Standard Model.
If this is the case, particle physics experiments
would provide complimentary data for such models. 
See \cite{Inami:2009bs,Furuuchi:2013ila}
for earlier investigations along this line 
in the case of extra-natural inflation.

\vskip10mm
\centerline{\bf Acknowledgments}\vskip2mm
\noindent KF benefited from the discussions 
on the BICEP2 results,
naturalness and WGC
at the Physical Research Laboratory (PRL) and 
Indian Institute of Astrophysics (IIA)
during his visits including the workshop
``Aspects of Cosmology" at IIA held in April 9-11, 2014.
In particular, he would like to express special thanks to
Namit Mahajan at PRL and Pravabati Chingangbam at IIA 
for the hospitality.
YK's work is supported in part 
by the National Science Council of Taiwan
under Grant No.~NSC-101-2112-M-007-021
and Taiwan String Theory Focus Group
of NCTS under Grant No.~NSC-103-2119-M-002-001.
The authors would also like to express their gratitude
to the anonymous referee for various suggestions for improvements,
in particular for pointing out possible relevance of the
effects of the axionic couplings in the current analysis.

\appendix

\section{Four-Dimensional Effective Action from \\
Higher-Dimensional Gauge Theory}\label{HDG}

\subsection{One-loop Effective Potential}\label{AppOne}

In this appendix we outline the calculation
of the one-loop effective potential
in higher-dimensional gauge theories
compactified on a circle.
We start with the five-dimensional action
\begin{align}
S=&\int d^5x\left[-\frac14 F^A_{MN}F^{A\,MN}-\frac14F^B_{MN}F^{B\,MN}
+\frac12 m_A^2(A_M-g_A\partial_M \theta)^2+{\bar \psi}i\Gamma^{M}D_M\psi \right]
\nn\\
&+S_{g.f.},
\end{align}
where space-time indices $M$ and $N$ run $0,\cdots,3$ and $5$, 
\begin{equation}
F^A_{MN}=\partial_M A_N-\partial_N A_M,\quad F^B_{MN}=\partial_M B_N-\partial_N B_M,
\end{equation}
and
\begin{equation}
D_M\psi=\partial_M\psi-ig_{A5}pA_M\psi-ig_{B5}qB_M\psi.
\end{equation}
We choose the gauge fixing term as
\begin{equation}
S_{g.f.}=\int d^5x
\left[-\frac12(\partial_MA^M)^2 -\frac12 (\partial_MB^M)^2 \right].
\end{equation}
Then the total action becomes 
\begin{equation}
S=\int d^5x
\Bigl[
\frac12 A_N\partial_M\partial^MA^{N}+\frac12B_N\partial_M\partial^MB^{N}
+\frac{m_A^2}{2}\left(A_M-g_A\pa_M\theta\right)^2
+{\bar \psi}i\Gamma^{M}D_M\psi
\Bigr].
\end{equation}
We compactify the fifth dimension on a circle with radius $L$. 
The Fourier expansions of the fields 
in the fifth dimension are
\begin{align}
A_M(x,x^5)&=\frac{1}{\sqrt{2\pi L}}
\sum^{\infty}_{n=-\infty}A_{M(n)}(x)e^{i\frac{n}{L}x^5},
\quad \mbox{similar for $B_M$, $\psi$},\\ 
\theta(x,x^5) &=  \frac{x^5}{g_5 L} w 
+ \frac{1}{\sqrt{2\pi L}}\sum_{n=-\infty}^{\infty} 
\theta_{(n)}(x) e^{i\frac{n}{L}x^5}.
\end{align}
We will be interested in the effective potential for
the zero-modes of the gauge fields, 
$A_{5(0)}\equiv A$ and $B_{5(0)}\equiv B$.
At one-loop level, only the quadratic part
of the matter action is relevant:
\begin{equation}
S^{(2)}_{\psi}=\int d^4x\sum_{n=-\infty}^{\infty}
{\bar \psi}_{(n)}\left(i\Gamma^{\mu}\partial_{\mu}
+g_{A}p\Gamma^5A_{5(0)}+g_{B}q\Gamma^5B_{5(0)}
+\Gamma^5\frac{n}{L}\right)\psi_{(n)}.
\end{equation}
Here, $\mu$ and $\nu$ run four-dimensional space-time indices $0,\cdots,3$.
Then, the one-loop effective potential is expressed as
\begin{align}
V(A,B)_{\rm 1-loop}&={\rm Tr}\, {\rm ln} \left(-i \Gamma^{\mu}_E\partial_{\mu E}
-g_{A}p\Gamma^5_E A_{5(0)}+g_{B}q\Gamma^5_E B_{5(0)}+\Gamma^5_E\frac{n}{L}\right)\notag\\
&=\frac12{\rm Tr}\, {\rm ln} {\mbox{1}\hspace{-0.25em}\mbox{l}}_{4 \times 4} \left[ -\partial^2_{\mu E}+
\left(\frac{n}{L}-\left(g_A p A + g_B q B\right)\right)^2\right],
\end{align}
where we have made Wick rotation and
the subscript $E$ indicates the Euclidean space.
The four-dimensional gauge couplings are related to the
five-dimensional ones as
\begin{equation}
g_A = \frac{g_{A5}}{\sqrt{2\pi L}},\quad
g_B = \frac{g_{B5}}{\sqrt{2\pi L}}.
\label{eq:}
\end{equation}
Employing the $\zeta$ function regularization, 
the effective potential becomes
\begin{equation}
V(A,B)_{\rm 1-loop}
=
\frac{3}{\pi^2 (2\pi L)^4}
\sum_{n=1}^{\infty}\frac{1}{n^5}\cos \left[n\left(\frac{pA}{f_A}+\frac{qB}{f_B}\right)\right],
\label{enpot}
\end{equation}
where
\begin{equation}
f_A = \frac{1}{g_A (2\pi L)}, \quad
f_B = \frac{1}{g_A (2\pi L)}.
\end{equation}
In (\ref{enpot}) we have dropped the constant part,
the fine tuning of which is the cosmological constant problem 
which we will not address in this paper.
Taking the leading term $n=1$ in 
(\ref{enpot})
together with the tree-level potential coming from the 
Stueckelberg mass term,
we arrive at the potential
\begin{equation}
V(A,B)\simeq
\frac{m_A^2}{2}
\left(A-2\pi f w\right)^2 +
\frac{3}{\pi^2 (2\pi L)^4}
\left[1-\cos\left(\frac{pA}{f_A}+\frac{qB}{f_B}\right) \right],
\label{apppot}
\end{equation}
where we have redefined the field $B$ by an appropriate constant shift.

\subsection{Axionic Couplings}\label{appAC}

The shift symmetry allows
the following axionic coupling
\begin{equation}
S_{AC} 
= 
\int d^4x \,
\frac{\alpha \sigma}{4f} F_{\mu\nu} \tilde{F}^{\mu\nu} ,
\label{AAC}
\end{equation}
where $\sigma$ is an axion and $\alpha$ is some constant.
In higher-dimensional gauge theory, 
the axionic coupling (\ref{AAC}) follows from the
Chern-Simons term in five-dimensional gauge theory \cite{Furuuchi:2013ila}:
\begin{equation}
S_{CS}
=
\frac{k}{24\pi^2}
\int
{\cal A} {\cal F}^2 ,
\label{CS}
\end{equation}
where
${\cal A} = {\cal A}_M dx^M$,
${\cal F} = d{\cal A} = \frac{1}{2} {\cal F}_{MN} dx^M dx^N$
and $k$ is an integer.
Quantum corrections to $k$ due to parity-violating charged matters
are one-loop exact and proportional to the
cubic powers of charges \cite{Bonetti:2013ela}.
As we assume charges to be $\Ord(1)$, we may expect $k \sim \Ord(1-10)$.
The $1$-form ${\cal A}_M dx^M$
is related to the canonically normalized gauge field $A_M$ 
in five dimensions as
\begin{equation}
A_M = \frac{1}{g_5} {\cal A}_M ,
\label{canA}
\end{equation}
where $g_5$ is the five-dimensional gauge coupling.
After integrating KK modes of the fifth direction
we obtain the axionic coupling (\ref{AAC})
with
\begin{equation}
\alpha = \frac{g_4^2 k}{4\pi^2} ,
\label{alpha}
\end{equation}
and
\begin{equation}
\sigma = \frac{{\cal A}_{5(0)}}{g_4}.
\label{sigma}
\end{equation}

\section{Weak Gravity Conjecture}\label{AppWGC}

Weak Gravity Conjecture (WGC)
\cite{ArkaniHamed:2006dz}
asserts
the existence of a state with 
charge and mass $(q,m)$ which satisfy
\begin{equation}
\frac{g q}{\sqrt{4\pi}} \geq \sqrt{G_N} m = \frac{m}{\sqrt{8\pi} M_P}.
\label{qm}
\end{equation}
(\ref{qm}) is estimated from requiring that
the Coulomb repulsive force 
is greater than the Newtonian attractive force 
so that extremal black holes can loose their charge by emitting such particles.
In this paper we assume the existence
of a particle with the smallest unit charge,
with respect to which all charges are integers.
Generalization is straightforward and
dose not change the result qualitatively,
unless one assumes highly exotic charge spectrum.
Then, the
Dirac monopole with unit magnetic charge has 
charge and mass
\begin{equation}
q_m = \frac{4\pi}{g},\quad
m_m \simeq \frac{4\pi \Lambda_{UV}}{g^2},
\label{mm}
\end{equation}
where $\Lambda_{UV}$ is a UV scale which regularizes 
the mass of the Dirac monopole.
Here, we used non-Abelian gauge-Higgs system as the UV completion
to estimate the mass of the Dirac monopole.
An important constraint for our study is obtained by
applying WGC 
the Dirac monopole:
\begin{equation}
\frac{4\pi}{g} \gtrsim  \frac{4\pi\Lambda_{UV}}{g^2} \frac{1}{\sqrt{2}M_P}.
\label{magwgc}
\end{equation}
It follows that
\begin{equation}
\Lambda_{UV} \lesssim \sqrt{2} g M_P.
\label{cutoff}
\end{equation}
This condition also follows by requiring that
the Dirac monopole with unit magnetic charge
is not a black hole \cite{ArkaniHamed:2006dz}.
Strictly speaking, one should take into account
the running of the couplings.
We assume that those runnings are not significant 
so that they do not alter our order of magnitude estimate. 
In order for the higher-dimensional gauge theory to be applicable,
the compactification scale should be
sufficiently below the UV cut-off scale:\footnote{%
More precisely we consider WGC in five dimensions
\cite{ArkaniHamed:2006dz}.
In this case electro-magnetic dual to the one-form gauge potential
is two-form gauge potential which couples to magnetic strings.
Then the analysis of the forces in three spacial dimensions 
transverse to the string is the same.}
\begin{equation}
\frac{1}{L} \ll {\sqrt{2}} g M_P.
\label{Lbound}
\end{equation}
In terms of the axion decay constant $f = 1 / (g 2\pi L)$,
\begin{equation}
2\pi f \ll \sqrt{2} {M_P}.
\label{fbound}
\end{equation}
Since the above argument is an order estimate,
in the main body we adopted slightly milder bound 
$2\pi f \lesssim M_P$.

\section{Relevant Inflation Models 
in Light of BICEP2}\label{AppCMB}

In this appendix we review the constraints
from CMB observations, in particular
the possible detection of primordial tensor perturbation by
BICEP2 \cite{Ade:2014xna},
on 
inflation models
which are relevant in this paper. 
The detection of the B-mode polarization by BICEP2
indicates large tensor-to-scalar ratio $r$. 
In this paper we adopt a conservative value $r = 0.16$
at the pivot scale $k = 0.05$ Mpc$^{-1}$ as a reference value,
considering the uncertainty in the foreground
\cite{Flauger:2014qra}
and 
the constraint from Planck 2013 \cite{Ade:2013uln,Audren:2014cea}.

\subsection{Chaotic Inflation with Quadratic Potential}\label{appchao}
Consider quadratic potential for the inflaton
\begin{equation}
V(\phi) = \frac{m^2}{2} \phi^2.
\label{potch}
\end{equation}
We assume canonical kinetic term for the inflaton $\phi$.
The slow-roll parameters are given by
\begin{align}
\epsilon(\phi) &= \frac{M_P^2}{2}\left( \frac{V'}{V} \right)^2 = \frac{2M_P^2}{\phi^2},
\label{eps}\\
\eta(\phi) &= {M_P^2} \frac{V''}{V} = \frac{2M_P^2}{\phi^2}.
\label{eta}
\end{align}
We will use suffix $\ast$ to indicate that it is the value
when the pivot scale exited the horizon.
The scalar spectral index is given by
\begin{equation}
n_s = 1- 6\epsilon_\ast + 2\eta_\ast.
\label{ns}
\end{equation}
Using (\ref{eps}) and (\ref{eta}) we obtain
\begin{equation}
n_s = 1 - 0.04 \times \frac{r}{0.16}.
\label{nsch}
\end{equation}
The scalar power spectrum and the tensor power spectrum are given as
\begin{align}
P_s &= \frac{V(\phi_\ast)}{24\pi^2M_P^4\epsilon_\ast}
= 2.2 \times 10^{-9},
\label{Ps}\\
P_t & = \frac{2 V(\phi_\ast)}{3\pi^2M_P^4},
\label{Pt}
\end{align}
where the last value in (\ref{Ps}) is the COBE normalization.
The tensor-to-scalar ratio $r$ is given by
\begin{equation}
r \equiv \frac{P_t}{P_s} = 16 \epsilon_\ast,
\label{reps}
\end{equation}
or equivalently
\begin{equation}
\epsilon_\ast = 0.01 \times \left( \frac{r}{0.16} \right).
\label{epsr}
\end{equation}
From (\ref{Ps}) this requires
\begin{equation}
V(\phi_\ast) \simeq
(2.0 \times10^{16}\, {\rm GeV})^4 
\times \left(\frac{r}{0.16} \right).
\label{V}
\end{equation}
Via the Friedmann equation
$V \simeq 3H^2 M_P^2$,
(\ref{V}) corresponds to the Hubble scale
\begin{equation}
H_\ast \simeq 1.0 \times 10^{14} 
\times \left( \frac{r}{0.16} \right)^{1/2}
\,{\rm GeV}.
\label{H}
\end{equation}
The slow-roll inflation ends when $\epsilon(\phi_e) \sim 1$.
This gives
\begin{equation}
\phi_e \sim \sqrt{2} M_P.
\label{end}
\end{equation}
The number of e-folds is given as
\begin{equation}
N_\ast = 
\left| \frac{1}{M_P^2}
\int_{\phi_{e}}^{\phi_\ast} 
d \phi \frac{V}{V'} 
\right|
=
\frac{1}{4M_P^2}
\left[\phi_\ast^2 -\phi_{e}^2 \right].
\label{N}
\end{equation}
Thus
\begin{equation}
\phi_\ast = 2M_P \sqrt{N_\ast - \frac{1}{2}}
\simeq 14 M_P \times 
\left(\frac{{N_\ast - \frac{1}{2}}}{50} \right)^{1/2}.
\label{phipiv}
\end{equation}
Putting this value to (\ref{potch}) and 
comparing it with (\ref{V}),
we obtain
\begin{equation}
m = \sqrt{\frac{2V_\ast}{\phi_\ast^2}}
= 3.4 \times 10^{13} \,{\rm GeV} 
\times
\left(\frac{50}{{N_\ast - \frac{1}{2}}} \right)^{1/2}
\times 
\left( \frac{r}{0.16} \right)^{1/2} .
\label{m}
\end{equation}

\subsection{Natural Inflation}\label{appnatural}

\begin{figure}[htbp]
\centering
\includegraphics[width=5in]{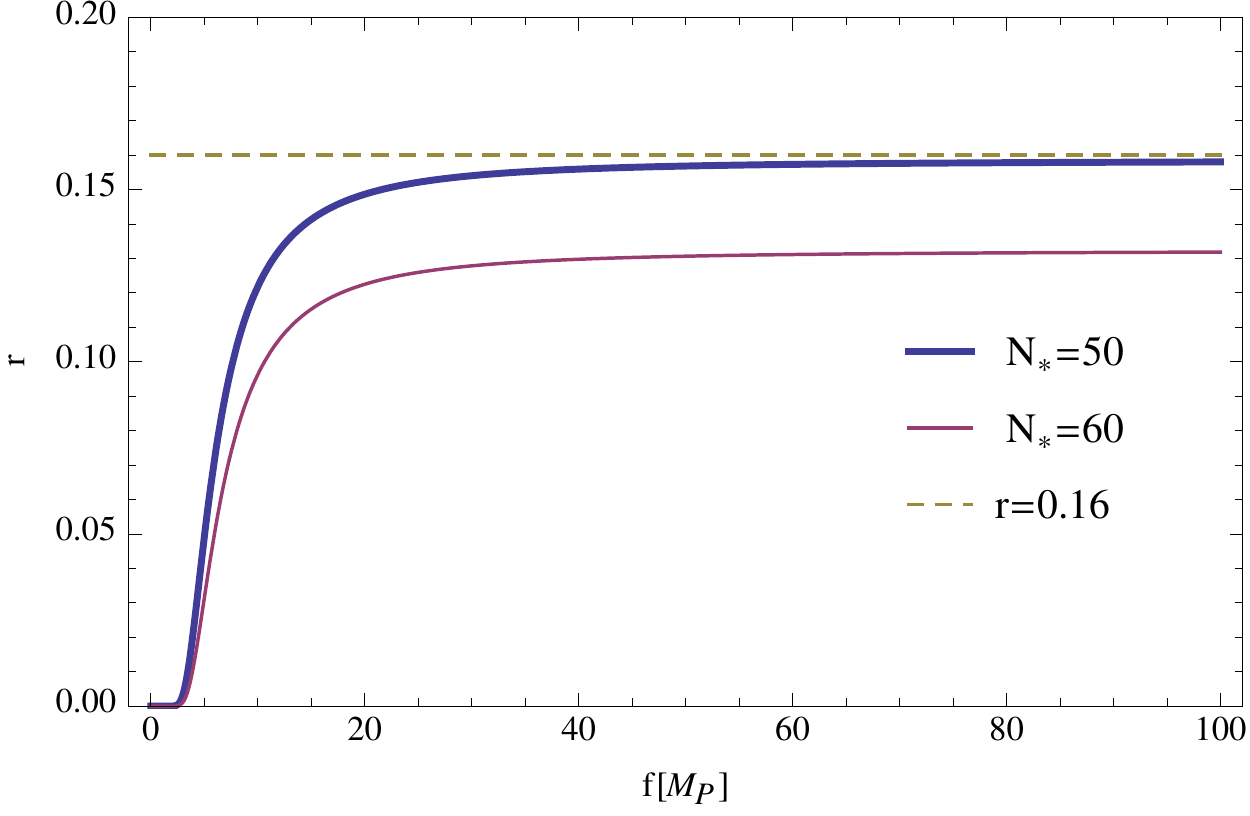} 
\caption{\label{fig:fr16} The tensor-to-scalar ratio
$r$ as a function of $f$ for two different values of $N_\ast$.}
\end{figure}

The typical form of the potential for 
natural inflation is given by
\begin{equation}
V(\phi) =
\frac{V_0}{2} \left[ 1 - \cos \left( \frac{\phi}{f} \right) \right] .
\label{iL}
\end{equation}
From (\ref{iL}) the slow-roll parameters are given as
\begin{align}
\epsilon (\phi)
&\equiv \frac{M_P^2}{2} \left( \frac{V'}{V} \right)^2 
= 
\frac{M_P^2}{2f^2}
\frac{ 1 + \cos \left( \frac{\phi}{f} \right)}{1 - \cos \left( \frac{\phi}{f} \right) } ,
\label{nepsilon} \\
\eta (\phi)
&\equiv M_P^2 \frac{V''}{V} 
= \frac{M_P^2}{f^2} \frac{\cos \left( \frac{\phi}{f} \right)}{1 - \cos \left( \frac{\phi}{f} \right)}.
\label{neta}
\end{align}
The number of e-folds as a function of $\phi$ is given by
\begin{align}
N(\phi)
&\simeq
\left|
\int_{\phi}^{\phi_e} d\phi\,
\frac{1}{M_P^2}\frac{V}{V'}
\right|
=
\left|
\int^{\phi}_{\phi_e} d\phi\,
\frac{f}{M_P^2}
\frac{1-\cos \frac{\phi}{f}}{\sin \frac{\phi}{f}}
\right|
\nn\\
&=
\left(
\frac{f}{M_P}
\right)^2
\left|
\log\left[\frac{1}{2}\left( 1 + \cos \frac{\phi}{f} \right)\right]_{\phi_e}^\phi 
\right|
\,.
\label{Nphi}
\end{align}
The slow-roll inflation ends when 
$\epsilon (\phi_e) \simeq 1$,
which gives
\begin{eqnarray}
\cos \frac{\phi_e}{f}
=
\left(
\frac{1-\frac{M_P^2}{2f^2}}{1+\frac{M_P^2}{2f^2}}
\right) .
\label{phie}
\end{eqnarray}
Plugging (\ref{phie}) into (\ref{Nphi}) we obtain
\begin{eqnarray}
\cos \frac{\phi}{f} =
\left(
\frac{2 e^{-\frac{M_P^2}{f^2}N}}{1 + \frac{M_P^2}{2f^2}} -1
\right) .
\label{phi}
\end{eqnarray}
From (\ref{nepsilon}) and (\ref{phi}),
for a given $N_\ast$, $r$ is determined as a function of $f$.
This is plotted in Fig.~\ref{fig:fr16}.
Notice that to obtain the tensor-to-scalar ratio
as large as $r\simeq 0.16$,
we need $f \gtrsim 20 M_P$ and $N_\ast \simeq 50$.
These values were adopted in the main body.

\bibliography{xdiref}

\providecommand{\href}[2]{#2}\begingroup\raggedright\begin{thebibliography}{10}

\bibitem{Guth:1980zm}
A.~H. Guth, {\it {The Inflationary Universe: A Possible Solution to the Horizon
  and Flatness Problems}},  {\em Phys.Rev.} {\bf D23} (1981) 347--356.

\bibitem{Linde:1981mu}
A.~D. Linde, {\it {A New Inflationary Universe Scenario: A Possible Solution of
  the Horizon, Flatness, Homogeneity, Isotropy and Primordial Monopole
  Problems}},  {\em Phys.Lett.} {\bf B108} (1982) 389--393.

\bibitem{Starobinsky:1980te}
A.~A. Starobinsky, {\it {A New Type of Isotropic Cosmological Models Without
  Singularity}},  {\em Phys.Lett.} {\bf B91} (1980) 99--102.

\bibitem{Sato:1980yn}
K.~Sato, {\it {First Order Phase Transition of a Vacuum and Expansion of the
  Universe}},  {\em Mon.Not.Roy.Astron.Soc.} {\bf 195} (1981) 467--479.

\bibitem{Kazanas:1980tx}
D.~Kazanas, {\it {Dynamics of the Universe and Spontaneous Symmetry Breaking}},
   {\em Astrophys.J.} {\bf 241} (1980) L59--L63.

\bibitem{Albrecht:1982wi}
A.~Albrecht and P.~J. Steinhardt, {\it {Cosmology for Grand Unified Theories
  with Radiatively Induced Symmetry Breaking}},  {\em Phys.Rev.Lett.} {\bf 48}
  (1982) 1220--1223.

\bibitem{Mukhanov:1981xt}
V.~F. Mukhanov and G.~V. Chibisov, {\it {Quantum Fluctuation and Nonsingular
  Universe. (In Russian)}},  {\em JETP Lett.} {\bf 33} (1981) 532--535.

\bibitem{Chibisov:1982nx}
G.~Chibisov and V.~F. Mukhanov, {\it {Galaxy formation and phonons}},  {\em
  Mon.Not.Roy.Astron.Soc.} {\bf 200} (1982) 535--550.

\bibitem{Guth:1982ec}
A.~H. Guth and S.~Pi, {\it {Fluctuations in the New Inflationary Universe}},
  {\em Phys.Rev.Lett.} {\bf 49} (1982) 1110--1113.

\bibitem{Hawking:1982cz}
S.~Hawking, {\it {The Development of Irregularities in a Single Bubble
  Inflationary Universe}},  {\em Phys.Lett.} {\bf B115} (1982) 295.

\bibitem{Starobinsky:1982ee}
A.~A. Starobinsky, {\it {Dynamics of Phase Transition in the New Inflationary
  Universe Scenario and Generation of Perturbations}},  {\em Phys.Lett.} {\bf
  B117} (1982) 175--178.

\bibitem{Ade:2013uln}
{\bf Planck Collaboration} Collaboration, P.~Ade et~al., {\it {Planck 2013
  results. XXII. Constraints on inflation}},
  \href{http://arxiv.org/abs/1303.5082}{{\tt arXiv:1303.5082}}.

\bibitem{Ade:2014xna}
{\bf BICEP2 Collaboration} Collaboration, P.~Ade et~al., {\it {Detection of
  B-Mode Polarization at Degree Angular Scales by BICEP2}},  {\em
  Phys.Rev.Lett.} {\bf 112} (2014) 241101,
  [\href{http://arxiv.org/abs/1403.3985}{{\tt arXiv:1403.3985}}].

\bibitem{Lyth:1996im}
D.~H. Lyth, {\it {What would we learn by detecting a gravitational wave signal
  in the cosmic microwave background anisotropy?}},  {\em Phys.Rev.Lett.} {\bf
  78} (1997) 1861--1863, [\href{http://arxiv.org/abs/hep-ph/9606387}{{\tt
  hep-ph/9606387}}].

\bibitem{Kim:2004rp}
J.~E. Kim, H.~P. Nilles, and M.~Peloso, {\it {Completing natural inflation}},
  {\em JCAP} {\bf 0501} (2005) 005,
  [\href{http://arxiv.org/abs/hep-ph/0409138}{{\tt hep-ph/0409138}}].

\bibitem{Silverstein:2008sg}
E.~Silverstein and A.~Westphal, {\it {Monodromy in the CMB: Gravity Waves and
  String Inflation}},  {\em Phys.Rev.} {\bf D78} (2008) 106003,
  [\href{http://arxiv.org/abs/0803.3085}{{\tt arXiv:0803.3085}}].

\bibitem{McAllister:2008hb}
L.~McAllister, E.~Silverstein, and A.~Westphal, {\it {Gravity Waves and Linear
  Inflation from Axion Monodromy}},  {\em Phys.Rev.} {\bf D82} (2010) 046003,
  [\href{http://arxiv.org/abs/0808.0706}{{\tt arXiv:0808.0706}}].

\bibitem{Berg:2009tg}
M.~Berg, E.~Pajer, and S.~Sjors, {\it {Dante's Inferno}},  {\em Phys.Rev.} {\bf
  D81} (2010) 103535, [\href{http://arxiv.org/abs/0912.1341}{{\tt
  arXiv:0912.1341}}].

\bibitem{Tye:2014tja}
S.~H.~H. Tye and S.~S.~C. Wong, {\it {Helical Inflation and Cosmic Strings}},
  \href{http://arxiv.org/abs/1404.6988}{{\tt arXiv:1404.6988}}.

\bibitem{Ben-Dayan:2014zsa}
I.~Ben-Dayan, F.~G. Pedro, and A.~Westphal, {\it {Hierarchical Axion
  Inflation}},  \href{http://arxiv.org/abs/1404.7773}{{\tt arXiv:1404.7773}}.

\bibitem{Kappl:2014lra}
R.~Kappl, S.~Krippendorf, and H.~P. Nilles, {\it {Aligned Natural Inflation:
  Monodromies of two Axions}},  {\em Phys.Lett.} {\bf B737} (2014) 124--128,
  [\href{http://arxiv.org/abs/1404.7127}{{\tt arXiv:1404.7127}}].

\bibitem{Long:2014dta}
C.~Long, L.~McAllister, and P.~McGuirk, {\it {Aligned Natural Inflation in
  String Theory}},  {\em Phys.Rev.} {\bf D90} (2014) 023501,
  [\href{http://arxiv.org/abs/1404.7852}{{\tt arXiv:1404.7852}}].

\bibitem{Bai:2014coa}
Y.~Bai and B.~A. Stefanek, {\it {Natural Milli-Charged Inflation}},
  \href{http://arxiv.org/abs/1405.6720}{{\tt arXiv:1405.6720}}.

\bibitem{Choi:2014rja}
K.~Choi, H.~Kim, and S.~Yun, {\it {Natural inflation with multiple
  sub-Planckian axions}},  {\em Phys.Rev.} {\bf D90} (2014) 023545,
  [\href{http://arxiv.org/abs/1404.6209}{{\tt arXiv:1404.6209}}].

\bibitem{Kaloper:2008fb}
N.~Kaloper and L.~Sorbo, {\it {A Natural Framework for Chaotic Inflation}},
  {\em Phys.Rev.Lett.} {\bf 102} (2009) 121301,
  [\href{http://arxiv.org/abs/0811.1989}{{\tt arXiv:0811.1989}}].

\bibitem{Kaloper:2011jz}
N.~Kaloper, A.~Lawrence, and L.~Sorbo, {\it {An Ignoble Approach to Large Field
  Inflation}},  {\em JCAP} {\bf 1103} (2011) 023,
  [\href{http://arxiv.org/abs/1101.0026}{{\tt arXiv:1101.0026}}].

\bibitem{Marchesano:2014mla}
F.~Marchesano, G.~Shiu, and A.~M. Uranga, {\it {F-term Axion Monodromy
  Inflation}},  {\em JHEP} {\bf 1409} (2014) 184,
  [\href{http://arxiv.org/abs/1404.3040}{{\tt arXiv:1404.3040}}].

\bibitem{Ashoorioon:2009wa}
A.~Ashoorioon, H.~Firouzjahi, and M.~Sheikh-Jabbari, {\it {M-flation: Inflation
  From Matrix Valued Scalar Fields}},  {\em JCAP} {\bf 0906} (2009) 018,
  [\href{http://arxiv.org/abs/0903.1481}{{\tt arXiv:0903.1481}}].

\bibitem{Harigaya:2014eta}
K.~Harigaya and M.~Ibe, {\it {Simple realization of inflaton potential on a
  Riemann surface}},  {\em Phys.Lett.} {\bf B738} (2014) 301--304,
  [\href{http://arxiv.org/abs/1404.3511}{{\tt arXiv:1404.3511}}].

\bibitem{McDonald:2014oza}
J.~McDonald, {\it {Sub-Planckian Two-Field Inflation Consistent with the Lyth
  Bound}},  {\em JCAP} {\bf 1409} (2014), no.~09 027,
  [\href{http://arxiv.org/abs/1404.4620}{{\tt arXiv:1404.4620}}].

\bibitem{Hosotani:1983xw}
Y.~Hosotani, {\it {Dynamical Mass Generation by Compact Extra Dimensions}},
  {\em Phys.Lett.} {\bf B126} (1983) 309.

\bibitem{Hatanaka:1998yp}
H.~Hatanaka, T.~Inami, and C.~Lim, {\it {The Gauge hierarchy problem and higher
  dimensional gauge theories}},  {\em Mod.Phys.Lett.} {\bf A13} (1998)
  2601--2612, [\href{http://arxiv.org/abs/hep-th/9805067}{{\tt
  hep-th/9805067}}].

\bibitem{ArkaniHamed:2003wu}
N.~Arkani-Hamed, H.-C. Cheng, P.~Creminelli, and L.~Randall, {\it {Extra
  natural inflation}},  {\em Phys.Rev.Lett.} {\bf 90} (2003) 221302,
  [\href{http://arxiv.org/abs/hep-th/0301218}{{\tt hep-th/0301218}}].

\bibitem{Kaplan:2003aj}
D.~E. Kaplan and N.~J. Weiner, {\it {Little inflatons and gauge inflation}},
  {\em JCAP} {\bf 0402} (2004) 005,
  [\href{http://arxiv.org/abs/hep-ph/0302014}{{\tt hep-ph/0302014}}].

\bibitem{Freese:1990rb}
K.~Freese, J.~A. Frieman, and A.~V. Olinto, {\it {Natural inflation with pseudo
  - Nambu-Goldstone bosons}},  {\em Phys.Rev.Lett.} {\bf 65} (1990) 3233--3236.

\bibitem{Banks:2003sx}
T.~Banks, M.~Dine, P.~J. Fox, and E.~Gorbatov, {\it {On the possibility of
  large axion decay constants}},  {\em JCAP} {\bf 0306} (2003) 001,
  [\href{http://arxiv.org/abs/hep-th/0303252}{{\tt hep-th/0303252}}].

\bibitem{ArkaniHamed:2006dz}
N.~Arkani-Hamed, L.~Motl, A.~Nicolis, and C.~Vafa, {\it {The String landscape,
  black holes and gravity as the weakest force}},  {\em JHEP} {\bf 0706} (2007)
  060, [\href{http://arxiv.org/abs/hep-th/0601001}{{\tt hep-th/0601001}}].

\bibitem{Banks:2010zn}
T.~Banks and N.~Seiberg, {\it {Symmetries and Strings in Field Theory and
  Gravity}},  {\em Phys.Rev.} {\bf D83} (2011) 084019,
  [\href{http://arxiv.org/abs/1011.5120}{{\tt arXiv:1011.5120}}].

\bibitem{Henneaux:1997ha}
M.~Henneaux and B.~Knaepen, {\it {All consistent interactions for exterior form
  gauge fields}},  {\em Phys.Rev.} {\bf D56} (1997) 6076--6080,
  [\href{http://arxiv.org/abs/hep-th/9706119}{{\tt hep-th/9706119}}].

\bibitem{Cheung:2014vva}
C.~Cheung and G.~N. Remmen, {\it {Naturalness and the Weak Gravity
  Conjecture}},  {\em Phys.Rev.Lett.} {\bf 113} (2014) 051601,
  [\href{http://arxiv.org/abs/1402.2287}{{\tt arXiv:1402.2287}}].

\bibitem{Flauger:2014qra}
R.~Flauger, J.~C. Hill, and D.~N. Spergel, {\it {Toward an Understanding of
  Foreground Emission in the BICEP2 Region}},  {\em JCAP} {\bf 1408} (2014)
  039, [\href{http://arxiv.org/abs/1405.7351}{{\tt arXiv:1405.7351}}].

\bibitem{Adam:2014bub}
{\bf Planck Collaboration} Collaboration, R.~Adam et~al., {\it {Planck
  intermediate results. XXX. The angular power spectrum of polarized dust
  emission at intermediate and high Galactic latitudes}},
  \href{http://arxiv.org/abs/1409.5738}{{\tt arXiv:1409.5738}}.

\bibitem{Audren:2014cea}
B.~Audren, D.~G. Figueroa, and T.~Tram, {\it {A note of clarification: BICEP2
  and Planck are not in tension}},  \href{http://arxiv.org/abs/1405.1390}{{\tt
  arXiv:1405.1390}}.

\bibitem{Trotta:2008qt}
R.~Trotta, {\it {Bayes in the sky: Bayesian inference and model selection in
  cosmology}},  {\em Contemp.Phys.} {\bf 49} (2008) 71--104,
  [\href{http://arxiv.org/abs/0803.4089}{{\tt arXiv:0803.4089}}].

\bibitem{Bayesian}
M.~P. Hobson~(Ed.), A.~H. Jaffe~(Ed.), A.~R. Liddle~(Ed.), P.~Mukherjee~(Ed.),
  and D.~Parkinson~(Ed.), {\it {Bayesian Methods in Cosmology}},  {\em
  Cambridge Univ. Press} (2009).

\bibitem{Shiu:2013wxa}
G.~Shiu, P.~Soler, and F.~Ye, {\it {Millicharged Dark Matter in Quantum Gravity
  and String Theory}},  {\em Phys.Rev.Lett.} {\bf 110} (2013), no.~24 241304,
  [\href{http://arxiv.org/abs/1302.5471}{{\tt arXiv:1302.5471}}].

\bibitem{McAllister:2014mpa}
L.~McAllister, E.~Silverstein, A.~Westphal, and T.~Wrase, {\it {The Powers of
  Monodromy}},  {\em JHEP} {\bf 1409} (2014) 123,
  [\href{http://arxiv.org/abs/1405.3652}{{\tt arXiv:1405.3652}}].

\bibitem{Flauger:2009ab}
R.~Flauger, L.~McAllister, E.~Pajer, A.~Westphal, and G.~Xu, {\it {Oscillations
  in the CMB from Axion Monodromy Inflation}},  {\em JCAP} {\bf 1006} (2010)
  009, [\href{http://arxiv.org/abs/0907.2916}{{\tt arXiv:0907.2916}}].

\bibitem{Meerburg:2014bpa}
P.~D. Meerburg, {\it {Alleviating the tension at low multipole through Axion
  Monodromy}},  \href{http://arxiv.org/abs/1406.3243}{{\tt arXiv:1406.3243}}.

\bibitem{'tHooft:1979bh}
G.~'t~Hooft, {\it {Naturalness, chiral symmetry, and spontaneous chiral
  symmetry breaking}},  {\em NATO Adv.Study Inst.Ser.B Phys.} {\bf 59} (1980)
  135.

\bibitem{Anber:2006xt}
M.~M. Anber and L.~Sorbo, {\it {N-flationary magnetic fields}},  {\em JCAP}
  {\bf 0610} (2006) 018, [\href{http://arxiv.org/abs/astro-ph/0606534}{{\tt
  astro-ph/0606534}}].

\bibitem{Barnaby:2011qe}
N.~Barnaby, E.~Pajer, and M.~Peloso, {\it {Gauge Field Production in Axion
  Inflation: Consequences for Monodromy, non-Gaussianity in the CMB, and
  Gravitational Waves at Interferometers}},  {\em Phys.Rev.} {\bf D85} (2012)
  023525, [\href{http://arxiv.org/abs/1110.3327}{{\tt arXiv:1110.3327}}].

\bibitem{Meerburg:2012id}
P.~D. Meerburg and E.~Pajer, {\it {Observational Constraints on Gauge Field
  Production in Axion Inflation}},  {\em JCAP} {\bf 1302} (2013) 017,
  [\href{http://arxiv.org/abs/1203.6076}{{\tt arXiv:1203.6076}}].

\bibitem{Linde:2012bt}
A.~Linde, S.~Mooij, and E.~Pajer, {\it {Gauge field production in supergravity
  inflation: Local non-Gaussianity and primordial black holes}},  {\em
  Phys.Rev.} {\bf D87} (2013), no.~10 103506,
  [\href{http://arxiv.org/abs/1212.1693}{{\tt arXiv:1212.1693}}].

\bibitem{Ferreira:2014zia}
R.~Z. Ferreira and M.~S. Sloth, {\it {Universal Constraints on Axions from
  Inflation}},  \href{http://arxiv.org/abs/1409.5799}{{\tt arXiv:1409.5799}}.

\bibitem{Furuuchi:2013ila}
K.~Furuuchi and J.~M. Wu, {\it {$U(1)_{B-L}$ extra-natural inflation with
  Standard Model on a brane}},  {\em Phys.Lett.} {\bf B729} (2014) 56--61,
  [\href{http://arxiv.org/abs/1310.4646}{{\tt arXiv:1310.4646}}].

\bibitem{Inami:2009bs}
T.~Inami, Y.~Koyama, C.~Lim, and S.~Minakami, {\it {Higgs-Inflaton Potential in
  5D Super Yang-Mills Theory}},  {\em Prog.Theor.Phys.} {\bf 122} (2009)
  543--551, [\href{http://arxiv.org/abs/0903.3637}{{\tt arXiv:0903.3637}}].

\bibitem{Bonetti:2013ela}
F.~Bonetti, T.~W. Grimm, and S.~Hohenegger, {\it {One-loop Chern-Simons terms
  in five dimensions}},  {\em JHEP} {\bf 1307} (2013) 043,
  [\href{http://arxiv.org/abs/1302.2918}{{\tt arXiv:1302.2918}}].

\end{thebibliography}\endgroup
\bibliographystyle{JHEP}
\end{document}